\documentclass[procedia]{easychair}

\usepackage{doc}
\usepackage{makeidx}

\def\procediaConference{99th Conference on Topics of Superb Significance (COOL 2014)}

\title{Novel Electronic Structures of Ru-Pnictides \\Ru$Pn$ ($Pn$ = P, As, Sb)}

\titlerunning{Novel Electronic Structures in Ru-Pnictides \ldots}

\author{
    H. Goto\inst{1}
\and
    T. Toriyama\inst{1}
\and
    T. Konishi\inst{2}
\and
    Y. Ohta\inst{1}\thanks{ohta@faculty.chiba-u.jp}
\\
}

\institute{
  Department of Physics, Chiba University, Chiba 263-8522, Japan\\
\and 
  Graduate School of Advanced Integration Science, Chiba Univeristy, Chiba 263-8522, Japan\\
}

\authorrunning{Goto, Toriyama, Konishi and Ohta}

\begin{document}

\maketitle


\begin{abstract}
Density-functional-theory-based electronic structure calculations are made to consider the 
novel electronic states of Ru-pnictides RuP and RuAs where the intriguing phase transitions 
and superconductivity under doping of Rh have been reported.  
We find that there appear nearly degenerate flat bands just at the Fermi level in the 
high-temperature metallic phase of RuP and RuAs; the flat-band states come mainly from 
the $4d_{xy}$ orbitals of Ru ions and the Rh doping shifts the Fermi level just above the 
flat bands.  The splitting of the flat bands caused by their electronic instability may then be 
responsible for the observed phase transition to the nonmagnetic insulating phase at low 
temperatures.  We also find that the band structure calculated for RuSb resembles that of 
the doped RuP and RuAs, which is consistent with experiment where superconductivity 
occurs in RuSb without Rh doping.  
\end{abstract}


%
%

\section{Introduction}
\label{sect:introduction}

\begin{figure}[tb]
\begin{centering}
\includegraphics[width=0.7\textwidth]{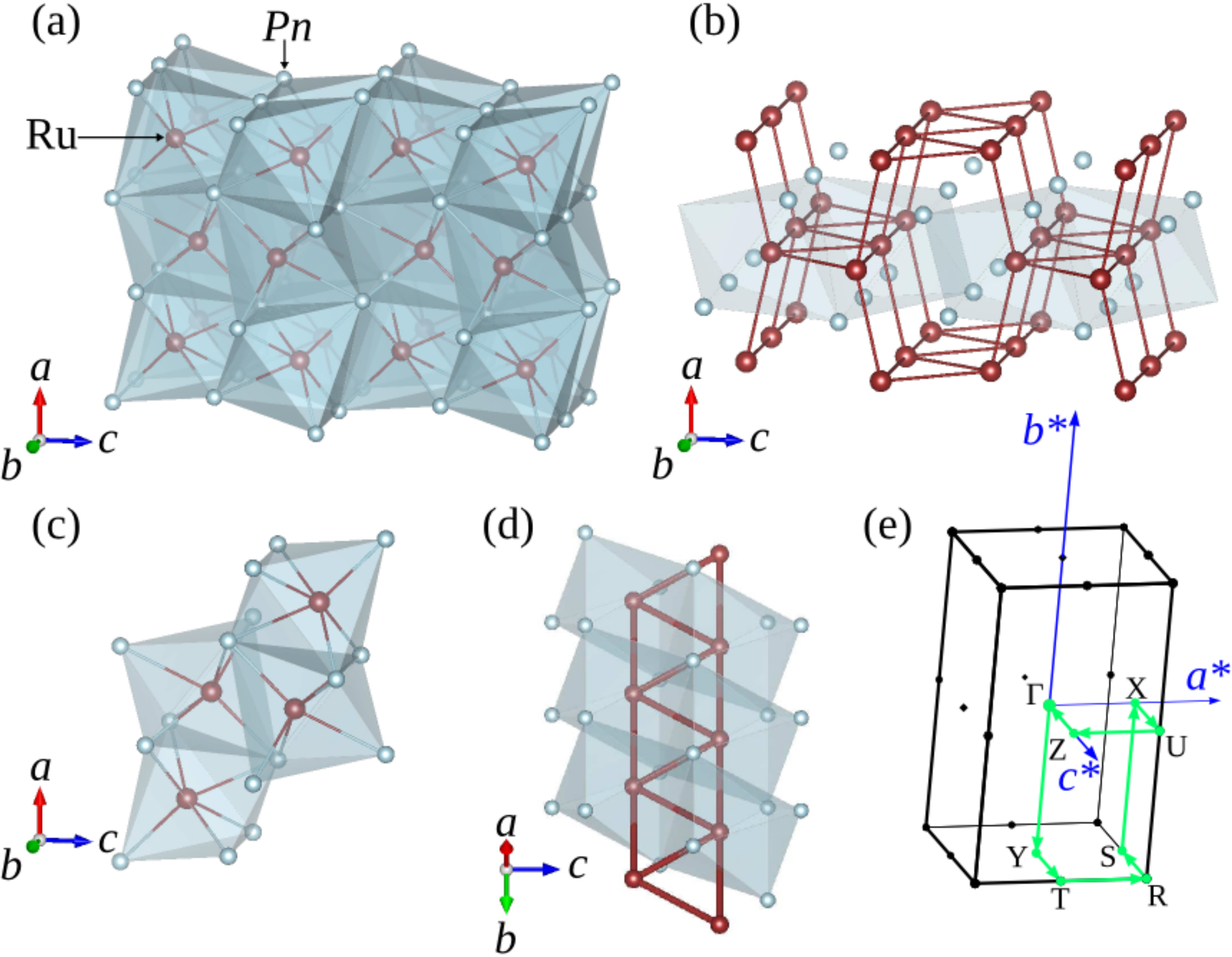}
\caption{
Schematic representations of the crystal structure of Ru$Pn$ ($Pn$ = P, As, Sb), 
where brown (gray) balls represent the Ru ($Pn$) ions.  
(a) The three-dimensional arrangement of the Ru$Pn_6$ octahedra, 
(b) the coupled zigzag chains of Ru ions, 
(c) the unit cell containing four Ru ions, 
(d) the zigzag ladder of Ru ions, and 
(e) the Brillouin zone of Ru$Pn$ in the high-temperature phase. 
}
\label{fig1}
\end{centering}
\end{figure}

Physics of transition-metal compounds has long been one of the major themes of strongly 
correlated electron systems \cite{khomskii}.  
Recently, superconductivity has been discovered in Ru-pnictides, RuP and RuAs, under Rh doping 
\cite{hirai,hirai2}.  These compounds crystallize in a MnP-type orthorhombic structure (space group 
$Pnma$), in which the Ru$Pn_6$ octahedra form a face-sharing chain along the $a$ axis 
\cite{rundqvist} of the crystal structure (see Fig.~\ref{fig1}).  The chains are connected by the edges 
and Ru ions form a distorted triangular lattice in the $bc$ plane.  The one-dimensional zigzag chains 
of the edge-sharing octahedra Ru$Pn_6$ are thus formed along the $b$ axis of the crystal.  

The undoped parent compounds RuP and RuAs show two sequential phase transitions: 
(i) a weak transition from a metal to a pseudogap phase at 330 K for RuP and at 280 K for RuAs, and 
(ii) a first-order transition to a nonmagnetic insulator phase at 270 K for RuP and at 200 K for RuAs.  
By Rh doping, these two phase transitions are suppressed and superconductivity occurs in 
the vicinity of the pseudogap phase \cite{hirai, hirai2}.  

The Ru $3d$ core-level and valence-band photoemission spectroscopy experiments on 
RuP \cite{sato} have shown that the Ru valence is $+3$ with $t_{2g}^5$ configuration and that the 
spectral weight near the Fermi level is moderately suppressed in the pseudogap phase, consistent 
with the pseudogap opening of $2\Delta/k_{\rm B}T_c\sim 3$ with the gap size $\Delta\sim 50$ meV 
and transition temperature $T_c\sim 330$ K.  It has also been suggested \cite{sato} that the 
electronic orderings responsible for the phase transitions are different from the conventional 
charge-density-wave ordering because the Ru $3d$ peak in the photoemission spectrum 
remains sharp in the pseudogap and insulating phases.  

The NMR experiment has also been done for Ru$_{1-x}$Rh$_x$P using $^{31}$P nuclei \cite{li}.  
It has been shown that, for the undoped RuP, both the Knight shift $K$ and relaxation rate 
$1/T_1T$ suddenly decrease at 270 K with decreasing temperature, indicating that the density 
of states at the Fermi level suddenly decreases at 270 K where the transition to the nonmagnetic 
insulator phase occurs.  The temperature dependence of the uniform magnetic susceptibility 
observed \cite{hirai} is consistent with this behavior.  The energy gap in the nonmagnetic 
insulating phase at $x=0$ is estimated to be 1218 K.  The Rh doping suppresses the changes 
in $K$ and $1/T_1T$ and the antiferromagnetic spin correlation is strongly enhanced in the pseudogap 
phase at $x=0.2$.  

Recently, RuP single crystals were synthesized, whereby the two structural phase transitions 
have been confirmed.  However, it was found that the resistivity drops monotonically upon 
temperature cooling below the second transition, indicating that the material shows metallic 
behavior in the lowest temperatures, which is in sharp contrast with the insulating ground state 
of polycrystalline samples \cite{chen}.  Optical conductivity measurements were also performed 
to reveal that a sudden reconstruction of the band structure over a broad energy scale and a 
significant removal of conducting carriers occur below the first phase transition, while a 
charge-density-wave-like energy gap opens below the second phase transition \cite{chen}.  
Thus, there is an essential conflict even in experimental situation; i.e., RuP at low temperatures 
is insulating in polycrystalline samples but metallic in single-crystalline samples.  
Possible off-stoichiometry of the single-crystalline samples may be the cause of this discrepancy 
and further experimental studies are now in progress \cite{hirai3}.  

In this paper, in order to elucidate the microscopic origins of the two phase transitions and consider 
the mechanism of superconductivity, we carry out the density-functional-theory (DFT) based 
electronic structure calculations using WIEN2k \cite{wien2k} and add a number of new aspects 
to the previous calculation \cite{hirai2}.  
The basic electronic structures of Ru-pnictides RuP, RuAs, and RuSb, including the effects of Rh 
doping, are thereby discussed, which will be the first step toward understanding of the novel 
electronic states of these materials.  

We thus find that there appear nearly degenerate flat bands just at the Fermi level in the metallic 
phase of RuP and RuAs and that these bands come mainly from the $4d_{xy}$ orbitals of Ru ions.  
We then suppose that these flat bands indicate an electronic instability, resulting in the structural 
transition, which may be responsible for the pseudogap formation (or metal-insulator transition) 
in these systems.  Here, possible occurrence of the spin-singlet formation may cause the nonmagnetic 
insulator phase at low temperatures.  
The effects of Rh doping on RuP and RuAs are also examined in the virtual crystal approximation 
and find that the rigid-band approximation works fairly well in these systems.  We also calculate 
the band structure of RuSb and show that this material is significantly different from RuP and RuAs 
in that its electronic structure without doping resembles that of the doped RuP and RuAs.  

\section{Computational details}
\label{sect:method}

We employ the WIEN2k code \cite{wien2k} based on the full-potential linearized 
augmented-plane-wave method and present the calculated results obtained in the 
generalized gradient approximation for electron correlations, where we use the 
exchange-correlation potential of Ref.~\cite{PBE96}.  The spin-orbit interaction is 
taken into account as indicated in the calculated results.  We use the crystal structures 
measured at room temperature \cite{rundqvist,heyding,endresen}, which have the 
orthorhombic symmetry (space group $Pnma$) with the lattice constants 
(in units of \AA) of 
$a=5.520$, $b=3.168$, and $c=6.120$ for RuP, 
$a=5.628$, $b=3.239$, and $c=6.184$ for RuAs, and 
$a=5.9608$, $b=3.7023$, and $c=6.5797$ for RuSb.  
The unit cell contains 4 Ru ions and 4 ligand ions, where all the Ru ions 
(and ligand ions) are crystallographically equivalent.  
See Fig.~\ref{fig1} for the sketches of the crystal structure.  

In the self-consistent calculations, we use 225, 300, and 270 ${\bf k}$-points 
in the irreducible part of the Brillouin zone for RuP, RuAs, and RuSb, respectively.  
We used the muffin-tin radii ($R_{\rm MT}$) of 
2.50 (Ru) and 1.95 (P) for RuP, 
2.32 (Ru) and 2.21 (As) for RuAs, and 
2.46 (Ru) and 2.46 (Sb) for RuSb in units of Bohr, and 
assume the plane-wave cutoff of $K_{\rm max}=7.00/R_{\rm MT}$.  
We use the codes VESTA \cite{momma} and XCrySDen \cite{kokalj} 
for graphical purposes.  

\begin{figure}[tb]
\begin{centering}
\includegraphics[width=0.6\textwidth]{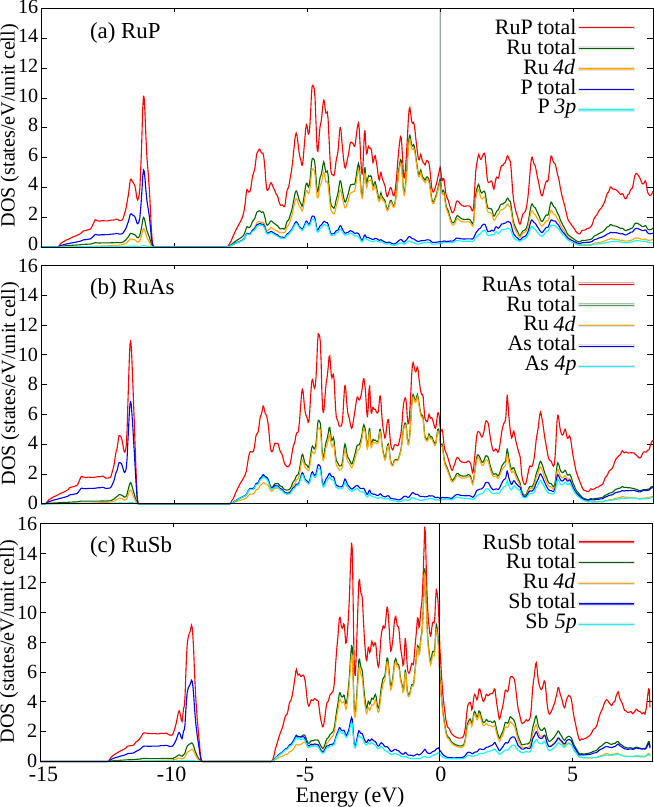}
\caption{
Calculated densities of states (DOS) of  (a) RuP, (b) RuAs, and (c) RuSb.  
The Fermi level is set to be the origin of energy and is indicated by the vertical line.  
The spin-orbit interaction is taken into account only for RuSb.  
}
\label{fig2}
\end{centering}
\end{figure}

\section{Results and discussion}
\label{sect:results}

\subsection{Densities of states}
\begin{figure}[tb]
\begin{centering}
\includegraphics[width=0.65\textwidth]{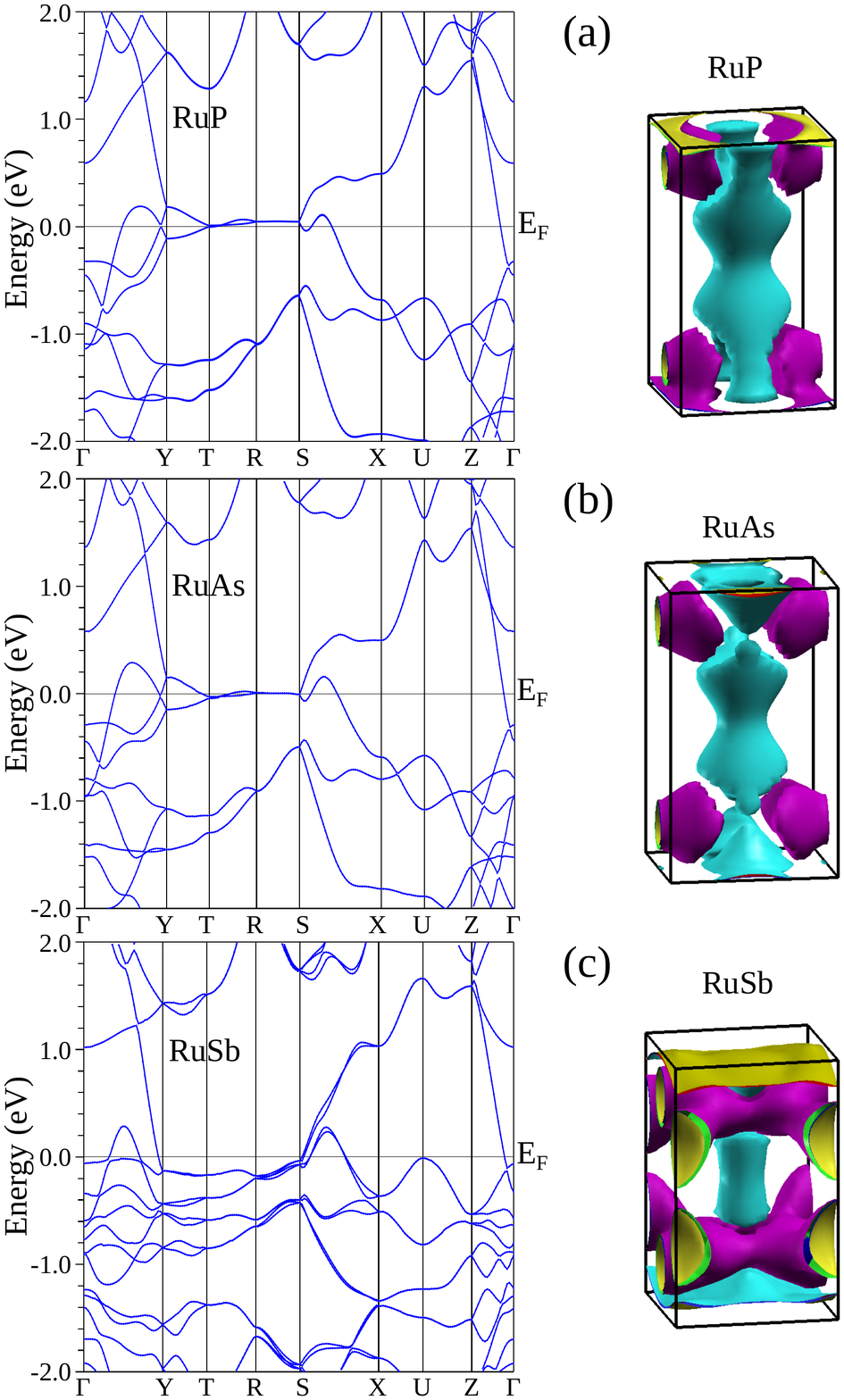}
\caption{
Calculated band dispersions and Fermi surfaces of (a) RuP, (b) RuAs, and (c) RuSb.  
The Fermi level is indicated by the horizontal line.  The spin-orbit interaction 
is taken into account only for RuSb.  
}
\label{fig3}
\end{centering}
\end{figure}

First, let us discuss the densities of states (DOS) of Ru$Pn$, which are shown in 
Fig.~\ref{fig2}.  We find that the topmost core $s$-orbital states of the ligand ions 
are located around $-12$ eV.  We then find that the $4d$ orbitals of Ru ions contribute 
to the DOS in a wide energy region between $-8$ eV and 5 eV and the states near 
the Fermi level are mainly come from the $4d$ $t_{2g}$ orbitals of Ru ions.  
The states coming from the $3p$ ($4p$) orbitals of P (As) ions are also 
extended in a wide energy range between $-8$ eV and $5$ eV, which are 
thus largely overlapped with the $4d$ states of Ru ions.  More precisely, 
the states of the $3p$ ($4p$) orbitals of P (As) ions are mainly located in 
the lower (between $-8$ and $-2$ eV) and higher (between $1.5$ and $5$ eV) 
energy regions and thus their contribution is rather small near the Fermi level.  
These results are in agreement with results of the previous calculation \cite{hirai2}.  
A sharp peak-like structure appears just at the Fermi level of the DOS of RuP and 
RuAs, which comes from the flat bands discussed in the next subsection.  
The situation of RuSb is rather different from that of RuP and RuAs as seen 
in Fig.~\ref{fig2}(c).  This occurs because the $5p$ states of Sb are located 
rather high in energy than those of the $3p$ ($4p$) states of P (As), as will 
be seen in Fig.~\ref{fig3}.  
Also noted is that the overall band width is rather narrower in RuSb than in 
RuP and RuAs, which is because the lattice constants of RuSb are considerably 
larger than those of RuP and RuAs, leading to a smaller hybridization and thus 
to the narrower band width in RuSb.  
The values of the DOS at the Fermi level are 5.33 for RuP, 5.09 for RuAs, and 
8.12 for RuSb in units of states/eV/unit cell.  

\subsection{Band dispersions and Fermi surfaces}
The calculated band dispersions and Fermi surfaces are shown in Fig.~\ref{fig3}.  
There are 32 bands in the energy range between $-8$ eV and $+5$ eV, 8 of which 
are from the $e_g$ orbitals of 4 Ru ions in the unit cell and other 24 of which are 
from the $t_{2g}$ orbitals of 4 Ru ions and $p$ orbitals of the 4 ligand ions.  
There are 44 valence electrons in the unit cell, such that the top 8 bands coming 
mainly from the $e_g$ orbitals of Ru are above the Fermi level, the next 4 bands 
coming mainly from  the $t_{2g}$ orbitals of Ru cross the Fermi level, and the lower 
20 bands coming from the $t_{2g}$ of Ru and $p$ orbitals of ligand ions are below 
the Fermi level.  

Near the Fermi level, we find four (or two doubly degenerate) bands consisting 
mainly from the $4d_{xy}$ orbitals of Ru, which form very flat bands just at the 
Fermi level in RuP and RuAs, in agreement with a previous calculation \cite{hirai2}.  
The flat-band states are located around the Y, T, R, and S points (or at the edge 
$k_y=\pm\pi/b$) of the Brillouin zone [see Fig.~\ref{fig1}(e) for the definition].  
The two bands of the four are mainly above the Fermi level and the other two band 
are mainly below the Fermi level, so that the semimetal-like Fermi surfaces 
of an equal number of electrons and holes are formed as shown in Fig.~\ref{fig3}.  
In RuSb, the situation is somewhat different; i.e., there are the flat bands (though 
rather deformed) but they are slightly below the Fermi level by $\sim$0.2 eV, 
indicating that a small number of electrons is transferred from the ligand ions.  
Thus, the Fermi surfaces of RuP and RuAs have the similar topology, which are 
however very different from the Fermi surface of RuSb.  
No clear nesting features are observed in the calculated Fermi surfaces of the 
present materials.  

The effects of Rh doping for Ru are also examined by the virtual crystal 
approximation.  We find that the doping shifts the Fermi level upward without 
any significant changes in the band structure; i.e., the rigid band approximation 
works well.  Thus, the effect of doping of Rh in RuP and RuAs results in 
the situation similar to the band structure of RuSb.  This result is consistent 
with the experimental fact that the superconductivity occurs by doping of Rh in RuP 
and RuAs, which occurs without Rh doping in RuSb \cite{hirai}.  

We also suggest that the degenerate flat bands at the Fermi level in RuP and 
RuAs should cause the electronic instability in these systems, the splitting of 
which may explain the insulating or pseudogap situation observed in these 
materials.  Further experimental and theoretical studies are required for clarifying 
the situations.  

\begin{figure}[tb]
\begin{centering}
\includegraphics[width=0.7\textwidth]{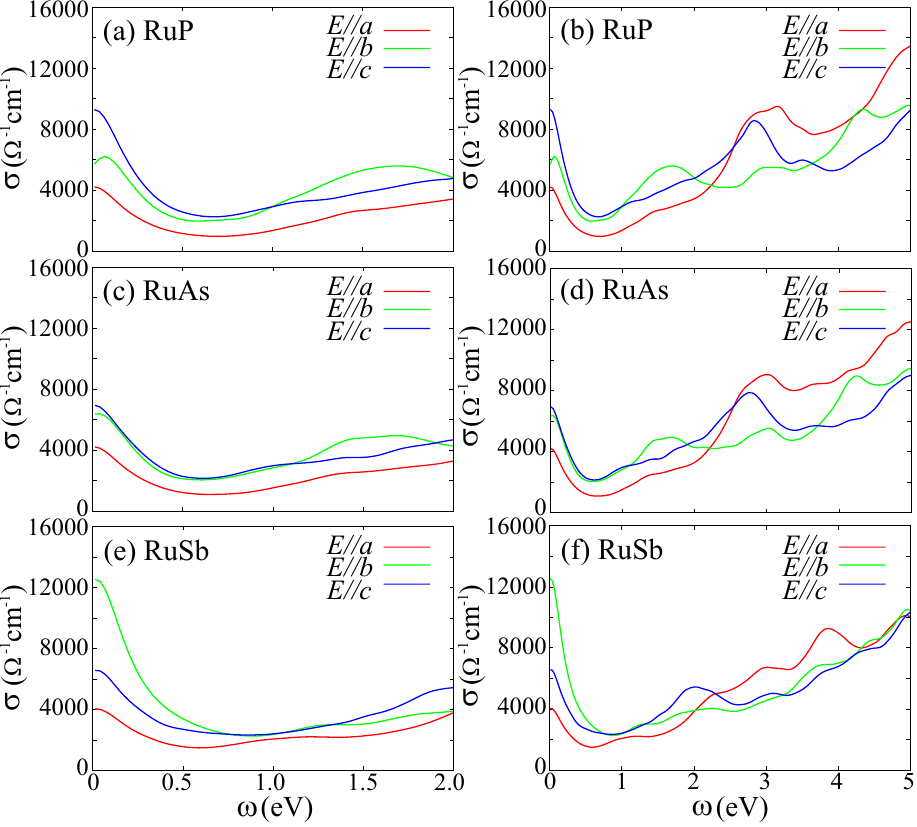}
\caption{
Calculated optical conductivity spectra $\sigma(\omega)$ of 
RuP [(a) and (b)], RuAs [(c) and (d)], and RuSb [(e) and (f)], where $E$ is the applied electric field.  
The spin-orbit interaction is taken into account.  Low-energy (high-energy) regions of the spectra 
are shown in the left (right) panels.  
}
\label{fig4}
\end{centering}
\end{figure}

\subsection{Optical conductivity}
We calculate the real part of the optical conductivity tensor $\Re\sigma_{\alpha\beta}(\omega)$ 
in the random-phase approximation using the results of our electronic structure calculations 
\cite{ambrosch}.  The results are shown in Fig.~\ref{fig4}, where both the interband and intraband 
(or Drude) contributions are included.  For the Drude contributions, we assume the form 
\begin{equation}
\Re\sigma_{\alpha\beta}(\omega)=\frac{\omega_{p,\alpha\beta}^2}{4\pi}\frac{\Gamma}{\omega^2+\Gamma^2},
\end{equation}
where $\omega_{p,\alpha\beta}$ is the plasma frequency calculated from the band structure 
and $\Gamma$ is a lifetime broadening ($\Gamma=0.10$ and 0.30 eV are assumed for the spectra 
of ${E\parallel a,b}$ and $E\parallel c$, respectively).  
The calculated values of the plasma frequency are 
2.502, 2.858, and 3.709 eV for the electric field along the $a$, $b$, and $c$ axes, respectively, for RuP; 
2.496, 3.056, and 3.204 eV for RuAs; and 
2.445, 4.305, and 3.113 eV for RuSb.  
The results for RuP can be compared with recent experimental data observed for a single-crystalline 
RuP \cite{chen}.  
We find that there is a double-peak structure in the calculated spectrum of 
$E\parallel b$ in a wide energy range up to $\sim$4 eV, which is in fair agreement  
with experiment.  A low-energy dip-like structure at $\sim$0.5 eV just above the Drude peak 
is also consistent with experiment.  The temperature variation of the spectra should be 
studied further to clarify the mechanism of the observed phase transitions.  

\begin{figure}[tb]
\begin{centering}
\includegraphics[width=0.8\textwidth]{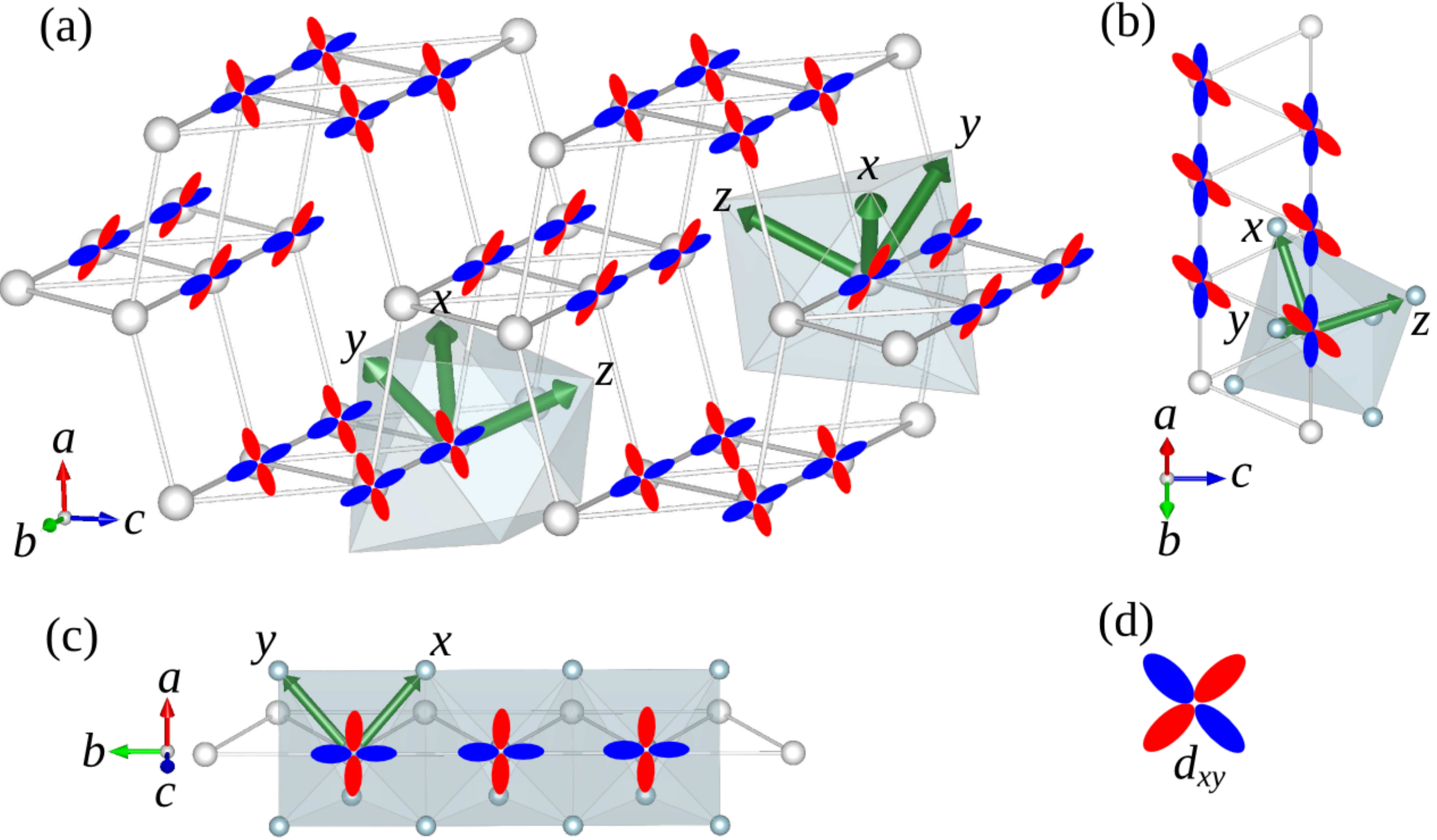}
\caption{
Arrangement of the $4d_{xy}$ orbitals of Ru ions in Ru$Pn$ forming the electronic states 
near the Fermi level.  
}
\label{fig5}
\end{centering}
\end{figure}

\subsection{States near the Fermi level}
To analyze the states near the Fermi level further, we calculate the orbital components 
of the states using the weight plots of the band dispersions and orbital-decomposed 
partial densities of states (although the detailed results are not shown here).  
We then find that the flat-band states at the Fermi level in RuP and RuAs are constructed 
mainly by the $4d_{xy}$ orbitals of Ru ions, which are illustrated schematically in Fig.~\ref{fig4}; 
i.e., the relevant $4d_{xy}$ orbitals are located on the one-dimensional zigzag ladder of Ru ions 
running along the $b$ direction of the crystal axis.  The zigzag ladders are then connected 
in the $ab$ plane to form the three-dimensional crystal structure.  
Thus, the low-energy effective model of the present material may be given by the 
connected zigzag ladders of the $4d_{xy}$ orbitals.  
We anticipate that the structural instability in RuP and RuAs may well reside in this 
effective model; for further analysis, experimental determination of the low-temperature 
crystal structures of RuP and RuAs is highly desirable.  

Also noted is that, if we assume the ionization state of Ru$^{3+}$, as was observed in 
the x-ray photoemission spectroscopy experiment \cite{sato}, then the filling of 5 electrons 
in the $4d$ ($t_{2g}$) orbitals of a Ru ion, so that the ligand ions are in the ionization state 
of $3-$ having $p^6$ electrons.  The location of the $p$ orbitals of the ligand ions are 
important in the present systems because the $p$ orbitals are overlapped largely with 
the $4d$ orbitals of Ru ions and thus the electrons can flow from the ligand ions to the 
$4d$ orbitals of Ru, resulting in the shift of the Fermi level.  This is in particular the case 
when we consider the results for RuSb.  

Further analyses of the effective model will be presented in future publications.  

\section{Summary}
\label{sect:summary}

Motivated by the recent discovery of the nonmagnetic insulating and pseudogap phases, 
together with superconductivity under Rh doping, in Ru-pnictides RuP, RuAs, and RuSb, 
we have carried out the DFT-based electronic structure calculations for these materials 
and have clarified their basic electronic structures responsible for the observed novel 
electronic properties.  
We have shown the following: 
(i) The $4d$-orbital states of Ru are hybridized strongly with the $p$-orbital states of 
the ligand ions in the entire energy range between $-8$ and 5 eV around the Fermi level.  
(ii) The states near the Fermi level are constructed mainly by the $4d_{xy}$ orbitals of 
Ru ions, which form the one-dimensional zigzag ladders connected in the MnP-type crystal 
lattice.  
(iii) There appear the fourfold nearly degenerate flat bands just at the Fermi level in RuP 
and RuAs, which are located $\sim$0.2 eV below the Fermi level in RuSb.  The flat bands 
come mainly from the $4d_{xy}$ orbitals of Ru ions.  
(iv) The doping of Rh ions in RuP and RuAs shifts the Fermi level just above the flat 
bands in a rigid-band--like manner, producing the similar electronic state near the Fermi 
level obtained for the undoped RuSb.   
(v) The calculated optical conductivity spectra of RuP are in fair agreement with 
experiment in the wide energy range.  

We hope that further experimental and theoretical studies will be done in future for 
clarifying the origins of the observed intriguing properties of these materials and 
elucidating the mechanism of superconductivity.  

\section{Acknowledgments}
We thank D. Hirai, M. Itoh, Y. Kobayashi, S. Li, and T. Mizokawa for providing us with useful 
information and discussions on Ru-pnictides.  This work was supported in part by Futaba 
Electronics Memorial Foundation and by a Grant-in-Aid for Scientific Research 
(No.~26400349) from JSPS of Japan.  T.T. acknowledges support 
from the JSPS Research Fellowship for Young Scientists.

\end{document}